\documentclass[10pt,twocolumn]{article}

\usepackage[margin=0.75in]{geometry}
\usepackage{times}
\usepackage{amsmath,amssymb,amsfonts}
\usepackage{graphicx}
\usepackage{multirow}
\usepackage{booktabs}
\usepackage{url}
\usepackage{cite}
\usepackage{caption}
\usepackage{xcolor}
\usepackage{xcolor}
\usepackage{pifont}
\usepackage{booktabs}
\usepackage{array}

\newcommand{\cmark}{\textcolor{green!60!black}{\ding{51}}}
\newcommand{\xmark}{\textcolor{red!75!black}{\ding{55}}}
\title{\bf Communication Strategy Selection for Multi-GPU\\
3D FDTD with Convolutional Perfectly Matched Boundary Layers}

\author{
Victory Obieke\\
Department of Mathematics\\
Oregon State University\\
Corvallis, OR, USA\\
\texttt{obiekev@oregonstate.edu}
}

\date{}

\begin{document}

\maketitle

\begin{abstract}
\textit{In this paper we describe a communication-strategy study for multi-GPU three-dimensional finite-difference time-domain computation with convolutional perfectly matched layer boundary conditions using CUDA. The metrics used to determine the most effective implementation include runtime, throughput in millions of output points per second, strong-scaling efficiency, CPML overhead, host-staged versus direct GPU-to-GPU exchange speedup, and enlarged-ghost speedup. On a single NVIDIA Quadro RTX 6000 GPU, the CPML implementation sustains
2,889--3,290 million output points per second with less than 1\% boundary-layer
overhead, providing the single-GPU baseline for the multi-GPU study. The results show that direct GPU-to-GPU peer exchange is the dominant
optimization with a 2.46--2.76$\times$
speedup over host-staged exchange, while enlarged ghost regions give only modest benefits because
the reduced communication frequency is partly offset by redundant computation
and additional memory traffic. On NVIDIA Quadro RTX 8000 GPUs, the implementation gives up to a
1.51$\times$ speedup on two GPUs for the tested strong-scaling cases, while four GPUs enable larger grids that approach or exceed single-GPU memory capacity.
}

\end{abstract}

\section{Introduction}

Finite-difference time-domain (FDTD) methods are widely used for wave
propagation, electromagnetics, seismic modeling, and computational physics
because they combine structured-grid simplicity with regular stencil updates
\cite{yee1966numerical,taflove2005computational,mcmechan2008migration}.  This structure is well
suited for GPUs, where many grid points can be updated in parallel and large
memory bandwidth can be exploited \cite{nickolls2008scalable,lindholm2008nvidia,
micikevicius20093d}.  However, practical three-dimensional simulations
often require large domains, high-order stencils, and absorbing boundary
layers, so the memory footprint can exceed the capacity of a single GPU
\cite{micikevicius20093d,roden2000convolution,taflove2005computational}.

A central challenge in multi-GPU stencil computation is the balance between
local computation and inter-device communication.  In the standard one-step
halo exchange, neighboring GPUs exchange ghost layers after every time step.
This method is simple and avoids redundant work, but it can become
communication dominated when the local subdomain per GPU is small
\cite{micikevicius20093d,kamil2006implicit,meng2009performance}.  Enlarged ghost regions
reduce the communication frequency by exchanging a wider halo and advancing
several local time steps before the next exchange.  This idea is related to
temporal blocking and ghost-zone optimization for stencil computations
\cite{kamil2006implicit,meng2009performance,zhang2023revisiting}.

Most idealized stencil benchmarks omit boundary treatments that are essential
in production FDTD solvers.  Perfectly matched layers and convolutional
perfectly matched layers (CPML) are widely used to reduce artificial
reflections at computational boundaries \cite{berenger1994perfectly,roden2000convolution,
taflove2005computational}.  CPML changes the performance balance because it introduces
auxiliary variables, boundary-layer updates, and additional memory traffic.
Therefore, the benefit of enlarged ghost regions must be evaluated in the
presence of realistic absorbing boundary costs.

The goal of this work is not to show that four GPUs are always faster than
one GPU. Instead, the goal is to determine which communication strategy is
most effective when a 3D FDTD+CPML solver is distributed across multiple
GPUs. The contribution of this work is an empirical communication-strategy study showing that, for high-order 3D FDTD+CPML on peer-connected GPUs, direct GPU-to-GPU exchange is the dominant optimization, while enlarged ghost regions provide only limited additional benefit. We first compare baseline decomposition layouts, and then use the selected
pencil-\(yz\) layout to study host-staged exchange, direct GPU-to-GPU peer
exchange, and enlarged ghost-region communication.
The results show that direct GPU-to-GPU peer exchange is the most
important optimization, enlarged ghost regions give only modest additional
speedup, and multi-GPU decomposition is most valuable when larger grid sizes
approach or exceed the memory capacity of a single GPU.

\section{Related Work}

GPU acceleration of high-order three-dimensional finite-difference stencils has
been studied extensively.  Micikevicius \cite{micikevicius20093d} used data-access redundancy as a key
metric for optimizing 3D finite-difference computation on CUDA GPUs and also
discussed multi-GPU extensions.  Communication-avoiding
and temporal-blocking methods reduce communication or memory traffic by
computing multiple time steps locally, but they introduce redundant work in
overlap or ghost regions \cite{kamil2006implicit, meng2009performance,
zhang2023revisiting}.  The CPML formulation of Roden and Gedney provides an
efficient implementation of complex-frequency-shifted PML for FDTD simulations
\cite{roden2000convolution}.  In contrast to interior-only stencil benchmarks, the present
work includes CPML boundary layers and measures their effect in multi-GPU runs.
Table~\ref{tab:novelty_positioning} summarizes the positioning of this work
relative to closely related FDTD, CPML, and GPU stencil-computation studies.
The novelty of the present work is not the introduction of a new FDTD scheme,
but the combined performance evaluation of decomposition layout, host-staged
communication, direct GPU-to-GPU peer exchange, and enlarged ghost-region
communication for a practical multi-GPU 3D FDTD solver with CPML boundary
layers.
\begin{table*}[t]
\centering
\caption{\bf Positioning of the present work relative to closely related GPU stencil and FDTD studies.}
\label{tab:novelty_positioning}
\renewcommand{\arraystretch}{1.25}
\setlength{\tabcolsep}{4pt}
\resizebox{\textwidth}{!}{%
\begin{tabular}{p{0.28\textwidth}cccccc}
\toprule
\textbf{Study or research direction} 
& \textbf{Multi-GPU} 
& \textbf{CPML/PML} 
& \textbf{High-order 3D FDTD} 
& \textbf{Decomposition comparison} 
& \textbf{Host vs. peer exchange} 
& \textbf{Enlarged ghost regions} \\
\midrule

Classical FDTD and PML formulations \cite{yee1966numerical,berenger1994perfectly,roden2000convolution}
& \xmark & \cmark & \xmark & \xmark & \xmark & \xmark \\

GPU stencil optimization and temporal blocking studies \cite{kamil2006implicit,meng2009performance,zhang2023revisiting}
& \cmark & \xmark & \cmark & \xmark & \xmark & \cmark \\

GPU 3D finite-difference / FDTD acceleration studies \cite{micikevicius20093d,nickolls2008scalable}
& \cmark & \xmark & \cmark & \xmark & \xmark & \xmark \\

CPML-based practical FDTD simulations \cite{roden2000convolution,taflove2005computational}
& \xmark & \cmark & \cmark & \xmark & \xmark & \xmark \\

\textbf{Present work}
& \cmark & \cmark & \cmark & \cmark & \cmark & \cmark \\

\bottomrule
\end{tabular}%
}
\end{table*}
\section{Model}

\subsection{First-Order Acoustic Model and Discrete Update}

In the interior of the domain, away from the absorbing layers, we use the first-order acoustic pressure--velocity system $p_t=-K\nabla\cdot\mathbf{v}$ and $\mathbf{v}_t=-(1/\rho)\nabla p$, where $p$ is pressure, $\mathbf{v}=(v_x,v_y,v_z)$ is particle velocity, $K$ is the bulk modulus, and $\rho$ is density. This system implies the scalar wave equation $p_{tt}=c^2\Delta p$, where $c^2=K/\rho$. For the performance experiments in this work, the coefficients are absorbed into the wave-update scaling parameter, so the update is written in normalized form.

Let $\delta_x^{(2r)}$, $\delta_y^{(2r)}$, and $\delta_z^{(2r)}$ denote centered finite-difference derivative operators of spatial order $2r$. The spatial derivatives in the acoustic system are approximated by centered finite-difference operators. We write $\delta_x^{(2r)}$, $\delta_y^{(2r)}$, and $\delta_z^{(2r)}$ for the numerical approximations of $\partial_x$, $\partial_y$, and $\partial_z$, respectively. The $x$-derivative is approximated by $(\delta_x^{(2r)}f)_{i,j,k}=(1/\Delta x)\sum_{\ell=1}^{r}a_\ell(f_{i+\ell,j,k}-f_{i-\ell,j,k})$. Similarly, the $y$- and $z$-derivatives are approximated by $(\delta_y^{(2r)}f)_{i,j,k}=(1/\Delta y)\sum_{\ell=1}^{r}a_\ell(f_{i,j+\ell,k}-f_{i,j-\ell,k})$ and $(\delta_z^{(2r)}f)_{i,j,k}=(1/\Delta z)\sum_{\ell=1}^{r}a_\ell(f_{i,j,k+\ell}-f_{i,j,k-\ell})$. Here, $r$ is the stencil radius and $a_\ell$ are the finite-difference weights. In the eighth-order experiments, $r=4$.

Without CPML terms, the staggered pressure--velocity update is written componentwise as $v_x^{n+1/2}=v_x^{n-1/2}-(\Delta t/\rho)\delta_x^{(2r)}p^n$, $v_y^{n+1/2}=v_y^{n-1/2}-(\Delta t/\rho)\delta_y^{(2r)}p^n$, and $v_z^{n+1/2}=v_z^{n-1/2}-(\Delta t/\rho)\delta_z^{(2r)}p^n$. The pressure is then updated by $p^{n+1}=p^n-K\Delta t(\delta_x^{(2r)}v_x^{n+1/2}+\delta_y^{(2r)}v_y^{n+1/2}+\delta_z^{(2r)}v_z^{n+1/2})$.

Thus one full time step consists of a pressure-to-velocity stencil update followed by a velocity-to-pressure stencil update. Since each subupdate uses a radius-$r$ stencil, numerical dependence can extend by $2r$ grid cells during one full time step. Therefore, if $s$ local time steps are taken between halo exchanges, the enlarged ghost depth is chosen as $g=2rs$. This choice provides enough neighboring data for the local subdomain to advance $s$ steps before the next communication event.

\subsection{CPML Boundary Layers}

The absorbing boundary layer follows the CFS/Roden--Gedney-style CPML construction \cite{roden2000convolution}. In the interior of the domain, the first-order acoustic FDTD system is updated using the usual finite-difference derivatives. Inside the CPML layer, however, each spatial derivative is replaced by a stretched derivative plus a recursive memory correction. Thus, a discrete derivative such as $\delta_x f$ is replaced by $(1/\kappa_x)\delta_x f+\psi_x$, where $f$ can be the pressure $p$ or one of the velocity components $v_x$, $v_y$, and $v_z$. The variable $\psi_x$ is the CPML memory variable associated with the $x$-direction derivative.

The memory variable is updated recursively by $\psi_x^{n+1}=b_x\psi_x^n+a_x\delta_x f^n$. This recursive update is the practical implementation of the convolutional correction in CPML. In other words, the CPML correction depends on the history of the spatial derivative, but the full history is not stored. Instead, the single memory variable $\psi_x$ carries the required history information efficiently from one time step to the next.

The CPML coefficients are computed from damping and stretching profiles inside
the absorbing layer.  Let $\rho\in[0,1]$ denote the normalized distance into
the CPML layer.  In the performance code, cubic grading is used, so $q=3$ and
$\rho^q=\rho^3$.  The damping profile is
$\sigma_x(\rho)=\sigma_{\max}\rho^3$, where
$\sigma_{\max}=-4\log(R)/(2L_{\mathrm{CPML}}h)$ and $R=10^{-8}$.  The
stretching profile is $\kappa_x(\rho)=1+(\kappa_{\max}-1)\rho^3$, with
$\kappa_{\max}=5$.  The complex-frequency-shift profile is
$\alpha_x(\rho)=\alpha_{\max}(1-\rho)$, where
$\alpha_{\max}=0.05\sigma_{\max}$.  Analogous profiles are used in the
$y$- and $z$-directions.

The recursive coefficients are $b_x=\exp[-(\sigma_x/\kappa_x+\alpha_x)\Delta t]$ and $a_x=\sigma_x(b_x-1)/[\kappa_x(\sigma_x+\kappa_x\alpha_x)]$. The coefficient $b_x$ controls the decay of the memory variable, while $a_x$ controls how strongly the current derivative $\delta_x f^n$ contributes to the memory variable. Outside the CPML layer, $\sigma_x=0$, $\kappa_x=1$, and the memory variable is zero, so the modified derivative $(1/\kappa_x)\delta_x f+\psi_x$ reduces to the ordinary finite-difference derivative $\delta_x f$. Therefore, the CPML modification does not change the interior FDTD update.

For the velocity update, the standard interior update $v_x^{n+1/2}=v_x^{n-1/2}-\Delta t\,\delta_x p^n$ is replaced inside the CPML by $v_x^{n+1/2}=v_x^{n-1/2}-\Delta t[(1/\kappa_x)\delta_x p^n+\psi_{v,x}^n]$. Similarly, $v_y^{n+1/2}=v_y^{n-1/2}-\Delta t[(1/\kappa_y)\delta_y p^n+\psi_{v,y}^n]$, and $v_z^{n+1/2}=v_z^{n-1/2}-\Delta t[(1/\kappa_z)\delta_z p^n+\psi_{v,z}^n]$. Here, $\psi_{v,x}$, $\psi_{v,y}$, and $\psi_{v,z}$ are the CPML memory variables associated with the pressure derivatives used in the velocity updates.

The pressure update is modified in the same way. In the interior, pressure is updated from the divergence of the velocity field. Inside the CPML layer, this update becomes $p^{n+1}=p^n-\Delta t[(1/\kappa_x)\delta_x v_x^{n+1/2}+(1/\kappa_y)\delta_y v_y^{n+1/2}+(1/\kappa_z)\delta_z v_z^{n+1/2}+\psi_{p,x}^{n+1/2}+\psi_{p,y}^{n+1/2}+\psi_{p,z}^{n+1/2}]$. The memory variables $\psi_{p,x}$, $\psi_{p,y}$, and $\psi_{p,z}$ correspond to the velocity derivatives used in the pressure update. Thus, each coordinate direction has its own damping profile, stretching coefficient, and memory correction.

The CPML update is applied only in the global absorbing boundary layer. At internal GPU interfaces, no CPML is used. Neighboring GPU subdomains communicate only through halo exchange, while the CPML memory variables are updated only near the physical outer boundary of the global computational domain. This separation is important because CPML represents an absorbing physical boundary treatment, whereas halo exchange is only a communication mechanism used to provide neighboring grid values across internal GPU subdomain interfaces.

\subsection{Multi-GPU Decomposition and Halo Exchange}

The three-dimensional grid is decomposed across four GPUs using three layouts:
slab-\(z\), block-\(xy\), and pencil-\(yz\), corresponding to
\(1\times1\times4\), \(2\times2\times1\), and \(1\times2\times2\),
respectively. In the slab decomposition, the domain is split only in the
\(z\)-direction. In the block decomposition, the domain is split in \(x\) and
\(y\). In the pencil-\(yz\) decomposition, the domain is split in \(y\) and
\(z\), while the full \(x\)-direction remains local to each GPU.

Each GPU stores its owned subdomain plus ghost cells in the decomposed
directions. Physical outer boundaries are treated by CPML, while internal GPU
interfaces are treated only by halo exchange. Thus, CPML is applied only at the
global outer boundary of the computational domain, not at internal subdomain
interfaces.

The implementation uses raw CUDA kernels through CuPy to pack halo data into
contiguous send buffers, direct CUDA peer copies for GPU-to-GPU exchange when
available, and raw CUDA kernels to unpack the received data into ghost cells.
The local FDTD+CPML update is then applied on each GPU.

For standard halo exchange, \(s=1\), ghost regions are exchanged every time
step. For enlarged ghost exchange, each GPU exchanges a deeper halo and then
advances \(s\) local steps before the next exchange. Since one full first-order
acoustic time step contains both a pressure-to-velocity update and a
velocity-to-pressure update, the enlarged ghost depth is chosen as \(g=2rs\).
This reduces communication frequency but increases redundant computation and
memory traffic in the enlarged ghost region.

\begin{figure}[t]
    \centering
    \includegraphics[width=\columnwidth, trim=50 20 50 20, clip]{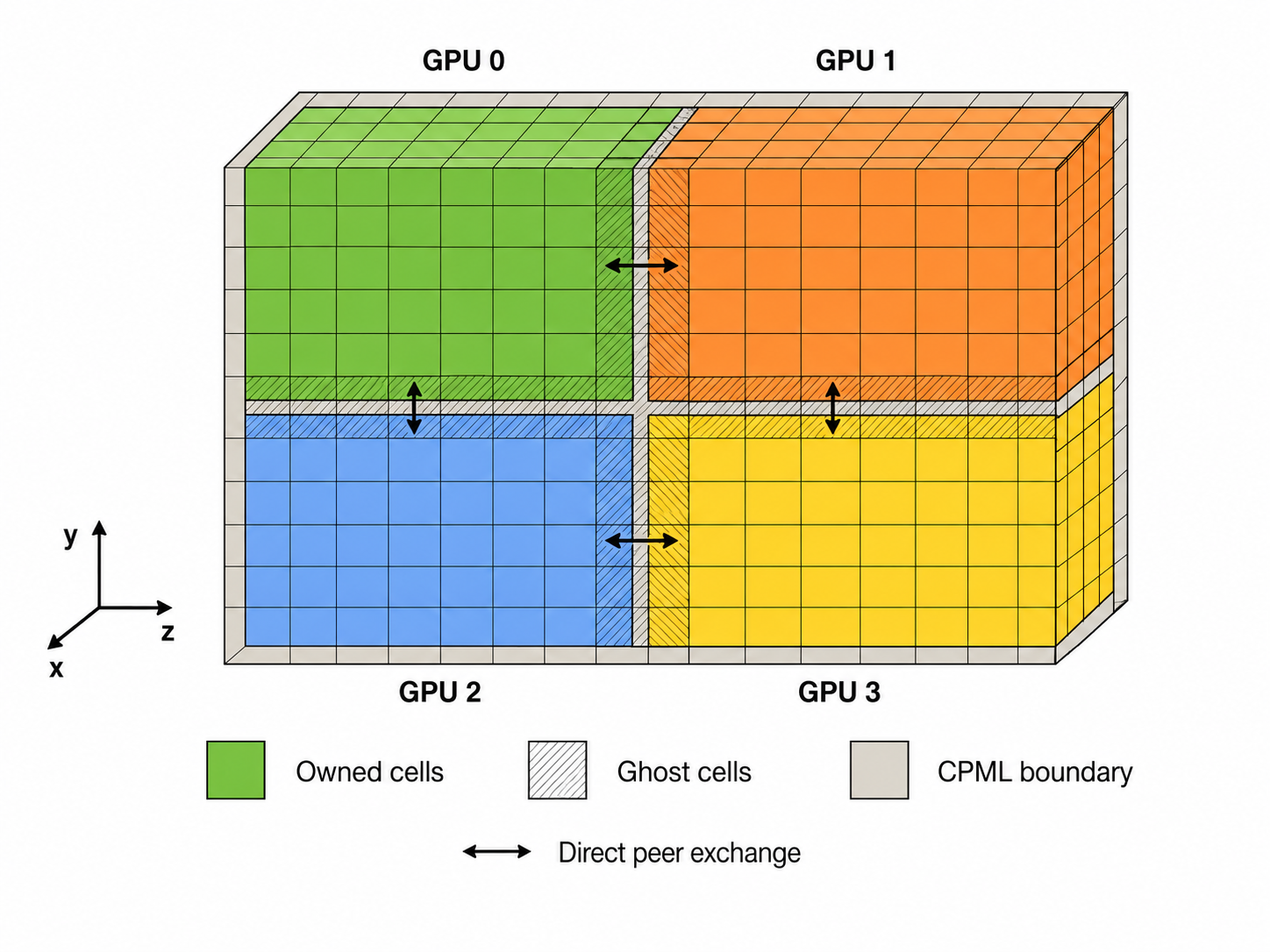}
    \caption{\bf Four-GPU pencil-\(yz\) decomposition.}
    \label{fig:gpuarch}
\end{figure}

\section{Experimental Setup}

Experiments use a raw-kernel implementation of the first-order acoustic
FDTD+CPML system. Unless otherwise stated, performance experiments use single
precision to emphasize communication and memory-bandwidth behavior. Reported
runtimes are medians over repeated runs.

\begin{table}[t]
\centering
\caption{\textbf{Numerical, CPML, and communication parameters used in the multi-GPU performance experiments.}}
\label{tab:experimental_parameters}
\renewcommand{\arraystretch}{1.15}
\setlength{\tabcolsep}{4pt}
\resizebox{\columnwidth}{!}{%
\begin{tabular}{|c|c|}
\hline
\textbf{Parameter} & \textbf{Value} \\
\hline
Governing system & First-order acoustic pressure--velocity FDTD system \\
\hline
Spatial stencil order & Eighth order \\
\hline
Stencil radius & $r=4$ \\
\hline
Time discretization & Two-stage pressure--velocity update \\
\hline
Wave-update scaling / CFL parameter & $\lambda_{\mathrm{FDTD}}=0.05$ \\
\hline
Communication intervals & $s\in\{1,2,4,8\}$ \\
\hline
Enlarged ghost depth & $g=2rs$ \\
\hline
Boundary condition & CFS/Roden--Gedney-style CPML \\
\hline
CPML application region & Global outer boundary only \\
\hline
Internal GPU interfaces & Halo exchange only, no CPML \\
\hline
CPML thickness & $L_{\mathrm{CPML}}=20$ cells \\
\hline
CPML target reflection parameter & $R=10^{-8}$ \\
\hline
CPML polynomial grading & Cubic grading, $q=3$ \\
\hline
CPML damping profile & $\sigma(\rho)=\sigma_{\max}\rho^3$ \\
\hline
CPML maximum damping & $\sigma_{\max}=-4\log(R)/(2L_{\mathrm{CPML}}h)$ \\
\hline
CPML stretching profile & $\kappa(\rho)=1+(\kappa_{\max}-1)\rho^3$ \\
\hline
CPML maximum stretching parameter & $\kappa_{\max}=5$ \\
\hline
CPML complex-frequency shift & $\alpha(\rho)=\alpha_{\max}(1-\rho)$ \\
\hline
CPML shift parameter & $\alpha_{\max}=0.05\sigma_{\max}$ \\
\hline
CPML auxiliary fields & Recursive memory variables for pressure and velocity derivatives \\
\hline
Performance precision & Single precision \\
\hline
Verification precision & Double precision where stated \\
\hline
Runtime statistic & Median over repeated runs \\
\hline
\end{tabular}
}
\end{table}

\subsection{Hardware and Software Environment}

The reported GPU experiments were run on the \texttt{optimus} node in the
\texttt{preempt} Slurm partition.  The node contains two Intel Xeon Gold 6230
CPUs and four NVIDIA Quadro RTX 6000 GPUs.  Because communication results
depend on the node topology, we record the CUDA software stack and the
GPU-to-GPU peer-access topology.

\begin{table}[t]
\centering
\caption{\bf Hardware and software environment for the main four-GPU RTX 6000 experiments.}
\label{tab:hardware_software_environment}
\renewcommand{\arraystretch}{1.15}
\setlength{\tabcolsep}{4pt}
\resizebox{\columnwidth}{!}{%
\begin{tabular}{|c|c|}
\hline
\bf Component & \bf Value \\
\hline
Node & \texttt{optimus.hpc.engr.oregonstate.edu} \\
\hline
Slurm partition & \texttt{preempt} \\
\hline
CPU & 2$\times$ Intel Xeon Gold 6230, 20 cores/socket \\
\hline
CPU cores & 40 physical cores, 2 NUMA nodes \\
\hline
System memory & 754 GiB \\
\hline
GPU model & NVIDIA Quadro RTX 6000 \\
\hline
Number of GPUs & 4 \\
\hline
GPU memory & 23040 MiB nominal; 22501 MiB available to CUDA per GPU \\
\hline
GPU driver & 590.48.01 \\
\hline
CUDA version reported by \texttt{nvidia-smi} & 13.1 \\
\hline
CUDA toolkit & 13.0, V13.0.88 \\
\hline
CUDA runtime / driver API & 13020 / 13010 \\
\hline
CuPy version & 14.1.0 \\
\hline
Python version & 3.10.14 \\
\hline
Operating system & Linux 5.14.0-570.58.1.el9\_6.x86\_64 \\
\hline
GPU topology & PCIe/NUMA paths with PIX, NODE, and SYS links; no NVLink reported \\
\hline
Peer access & Enabled for all ordered GPU pairs \\
\hline
Communication backend & CuPy CUDA copies with peer access enabled \\
\hline
NCCL & Not used \\
\hline
\end{tabular}
}
\end{table}

\begin{table}[t]
\centering
\caption{\bf CUDA peer-access matrix and topology labels for the selected four-GPU node. A value of 1 indicates that CUDA peer access is available. Topology labels are taken from \texttt{nvidia-smi topo -m}.}
\label{tab:peer_access_topology}
\renewcommand{\arraystretch}{1.15}
\setlength{\tabcolsep}{4pt}
\resizebox{\columnwidth}{!}{%
\begin{tabular}{|c|c|c|c|c|}
\hline
\bf From/To & \bf GPU 0 & \bf GPU 1 & \bf GPU 2 & \bf GPU 3 \\
\hline
\bf GPU 0 & -- & 1 / NODE & 1 / NODE & 1 / SYS \\
\hline
\bf GPU 1 & 1 / NODE & -- & 1 / PIX & 1 / SYS \\
\hline
\bf GPU 2 & 1 / NODE & 1 / PIX & -- & 1 / SYS \\
\hline
\bf GPU 3 & 1 / SYS & 1 / SYS & 1 / SYS & -- \\
\hline
\end{tabular}
}
\end{table}

In this topology, \texttt{PIX} denotes communication through at most one PCIe
bridge, \texttt{NODE} denotes communication through PCIe and host bridges
within a NUMA node, and \texttt{SYS} denotes communication that traverses PCIe
and the inter-socket CPU interconnect.  The peer-exchange results should
therefore be interpreted as CUDA peer-access results on a PCIe/NUMA-connected
four-GPU node, not as NVLink results.

Before applying the final raw-kernel communication optimizations, we first compare
the three four-GPU decomposition layouts. This baseline comparison is used only
to select the most favorable domain decomposition for the subsequent optimized
communication experiments.

\begin{table}[t]
\centering
\caption{\bf Baseline four-GPU decomposition comparison for the first-order acoustic 3D FDTD+CPML benchmark. This table uses the pre-optimized communication implementation.}
\label{tab:decomposition_comparison}
\renewcommand{\arraystretch}{1.15}
\setlength{\tabcolsep}{3pt}
\resizebox{\columnwidth}{!}{%
\begin{tabular}{|c|c|c|c|}
\hline
\bf Grid &
\bf slab-$z$ $(1\times1\times4)$ &
\bf block-$xy$ $(2\times2\times1)$ &
\bf pencil-$yz$ $(1\times2\times2)$ \\
\hline
$320^3$ &
2.357 s / 889.71 &
3.417 s / 613.72 &
\bf 2.345 s / 894.45 \\
\hline
$544^3$ &
8.399 s / 1226.79 &
10.889 s / 946.24 &
\bf 7.616 s / 1352.80 \\
\hline
$800^3$ &
23.778 s / 1378.06 &
29.947 s / 1094.19 &
\bf 21.932 s / 1494.09 \\
\hline
\end{tabular}%
}
\end{table}

Table~\ref{tab:decomposition_comparison} shows that the pencil-\(yz\)
decomposition gives the best throughput for all tested grids in the baseline
implementation. Therefore, the pencil-\(yz\) layout shown in
Figure~\ref{fig:gpuarch} is used for the remaining optimized four-GPU experiments.

\section{FDTD and CPML Implementation Verification}
\subsection{FDTD Verification}
To verify the finite-difference implementation, we use the periodic standing
wave \(u(x,y,z,t)=\sin(2\pi m x)\sin(2\pi m y)\sin(2\pi m z)
\cos(2\pi m\sqrt{3}\,ct)\), which satisfies \(u_{tt}=c^2\Delta u\) using the method of manufactured solutions

\begin{table}[t]
\centering
\caption{\bf Spatial convergence of the four-GPU pencil-\(yz\) finite-difference Laplacian for the periodic standing-wave solution with mode number \(m=4\).}
\label{tab:spatial_convergence_verification}
\renewcommand{\arraystretch}{1.15}
\setlength{\tabcolsep}{4pt}
\resizebox{\columnwidth}{!}{%
\begin{tabular}{|c|c|c|c|c|c|}
\hline
\bf Stencil order & \bf $N$ & \bf Relative $L^2$ error & \bf Rate & \bf Relative $L^\infty$ error & \bf Rate \\
\hline
\multirow{5}{*}{2}
& 30  & $5.714\times 10^{-2}$ & --   & $5.714\times 10^{-2}$ & --   \\ \cline{2-6}
& 60  & $1.454\times 10^{-2}$ & 1.97 & $1.454\times 10^{-2}$ & 1.97 \\ \cline{2-6}
& 120 & $3.650\times 10^{-3}$ & 1.99 & $3.650\times 10^{-3}$ & 1.99 \\ \cline{2-6}
& 240 & $9.135\times 10^{-4}$ & 2.00 & $9.135\times 10^{-4}$ & 2.00 \\ \cline{2-6}
& 480 & $2.284\times 10^{-4}$ & 2.00 & $2.284\times 10^{-4}$ & 2.00 \\
\hline
\multirow{5}{*}{4}
& 30  & $5.141\times 10^{-3}$ & --   & $5.141\times 10^{-3}$ & --   \\ \cline{2-6}
& 60  & $3.368\times 10^{-4}$ & 3.93 & $3.368\times 10^{-4}$ & 3.93 \\ \cline{2-6}
& 120 & $2.130\times 10^{-5}$ & 3.98 & $2.130\times 10^{-5}$ & 3.98 \\ \cline{2-6}
& 240 & $1.335\times 10^{-6}$ & 4.00 & $1.335\times 10^{-6}$ & 4.00 \\ \cline{2-6}
& 480 & $8.349\times 10^{-8}$ & 4.00 & $8.349\times 10^{-8}$ & 4.00 \\
\hline
\multirow{4}{*}{8}
& 30  & $6.568\times 10^{-5}$ & --   & $6.568\times 10^{-5}$ & --   \\ \cline{2-6}
& 60  & $2.891\times 10^{-7}$ & 7.83 & $2.891\times 10^{-7}$ & 7.83 \\ \cline{2-6}
& 120 & $1.164\times 10^{-9}$ & 7.96 & $1.164\times 10^{-9}$ & 7.96 \\ \cline{2-6}
& 240 & $4.561\times 10^{-12}$ & 7.99 & $4.606\times 10^{-12}$ & 7.98 \\
\hline
\end{tabular}
}
\end{table}

\begin{table}[t]
\centering
\caption{\bf Time convergence of the four-GPU pencil-\(yz\) wave solver using the eighth-order spatial stencil and mode number (m=1).}
\label{tab:time_convergence_verification}
\renewcommand{\arraystretch}{1.15}
\setlength{\tabcolsep}{5pt}
\resizebox{\columnwidth}{!}{%
\begin{tabular}{|c|c|c|c|c|}
\hline
\bf $\Delta t$ & \bf Relative $L^2$ error & \bf Rate & \bf Relative $L^\infty$ error & \bf Rate \\
\hline
$1.0\times 10^{-3}$ & $1.559\times 10^{-5}$ & --   & $1.559\times 10^{-5}$ & --   \\
\hline
$5.0\times 10^{-4}$ & $3.887\times 10^{-6}$ & 2.00 & $3.887\times 10^{-6}$ & 2.00 \\
\hline
$2.5\times 10^{-4}$ & $9.704\times 10^{-7}$ & 2.00 & $9.704\times 10^{-7}$ & 2.00 \\
\hline
\end{tabular}
}
\end{table}

Tables~\ref{tab:spatial_convergence_verification} and
\ref{tab:time_convergence_verification} show the expected spatial and temporal
orders of accuracy.
\subsection{CPML Verification}
We further verify the CPML implementation in Table~\ref{tab:cpml_reflection_decomposition_validation}.
\begin{table}[t]
\centering
\caption{\bf Reflected-to-incident amplitude ratio for different four-GPU decompositions.}
\label{tab:cpml_reflection_decomposition_validation}
\renewcommand{\arraystretch}{1.15}
\setlength{\tabcolsep}{5pt}
\resizebox{\columnwidth}{!}{%
\begin{tabular}{|c|c|c|c|}
\hline
\bf CPML cells
& \bf slab-$z$ $(1\times1\times4)$
& \bf block-$xy$ $(2\times2\times1)$
& \bf pencil-$yz$ $(1\times2\times2)$ \\
\hline
0  & $9.438\times10^{-1}$ & $9.438\times10^{-1}$ & $9.438\times10^{-1}$ \\
\hline
8  & $8.049\times10^{-2}$ & $8.049\times10^{-2}$ & $8.049\times10^{-2}$ \\
\hline
12 & $6.714\times10^{-3}$ & $6.715\times10^{-3}$ & $6.714\times10^{-3}$ \\
\hline
16 & $5.581\times10^{-4}$ & $5.581\times10^{-4}$ & $5.581\times10^{-4}$ \\
\hline
24 & $ \bf 3.328\times10^{-5}$ & $ \bf 5.300\times10^{-5}$ & $ \bf 3.328\times10^{-5}$ \\
\hline
\end{tabular}
}
\end{table}

Table~\ref{tab:cpml_reflection_decomposition_validation} confirms that the CPML
reduces artificial reflections by several orders of magnitude and behaves
consistently across decompositions. The reduction happens with more layers as expected in \cite{roden2000convolution}.

To complement the reflection-ratio verification, we also include a qualitative
wavefield visualization of the absorbing behavior of CPML.  This experiment
follows the acoustic CPML validation style of Pasalic and McGarry
\cite{pasalic2010convolutional}, who derive CPML for isotropic and anisotropic acoustic
wave equations and demonstrate its effectiveness using source-driven acoustic
wave propagation.  The purpose of the present visualization is not to reproduce
their full TTI anisotropic model exactly, but to provide a comparable acoustic
CPML demonstration for the pressure--velocity FDTD setting used in this work.

The visualization uses a two-dimensional acoustic-style pressure--velocity
system on a \(2\,{\rm km}\times 2\,{\rm km}\) domain with \(\Delta x=\Delta z=8\)
m.  Source \(s(t)=(1-2a^2)e^{-a^2}\), with
\(a=\pi f_0(t-t_0)\), is placed at the center of the domain with peak frequency
\(f_0=30\) Hz.  We set \(c_x=2600\) m/s and \(c_z=1800\) m/s.
The CPML thickness is \(10\) grid cells.  The same source, grid, time step, and
final time are used for the no-CPML and CPML simulations.

\begin{figure}[t]
    \centering
    \includegraphics[width=0.98\columnwidth]{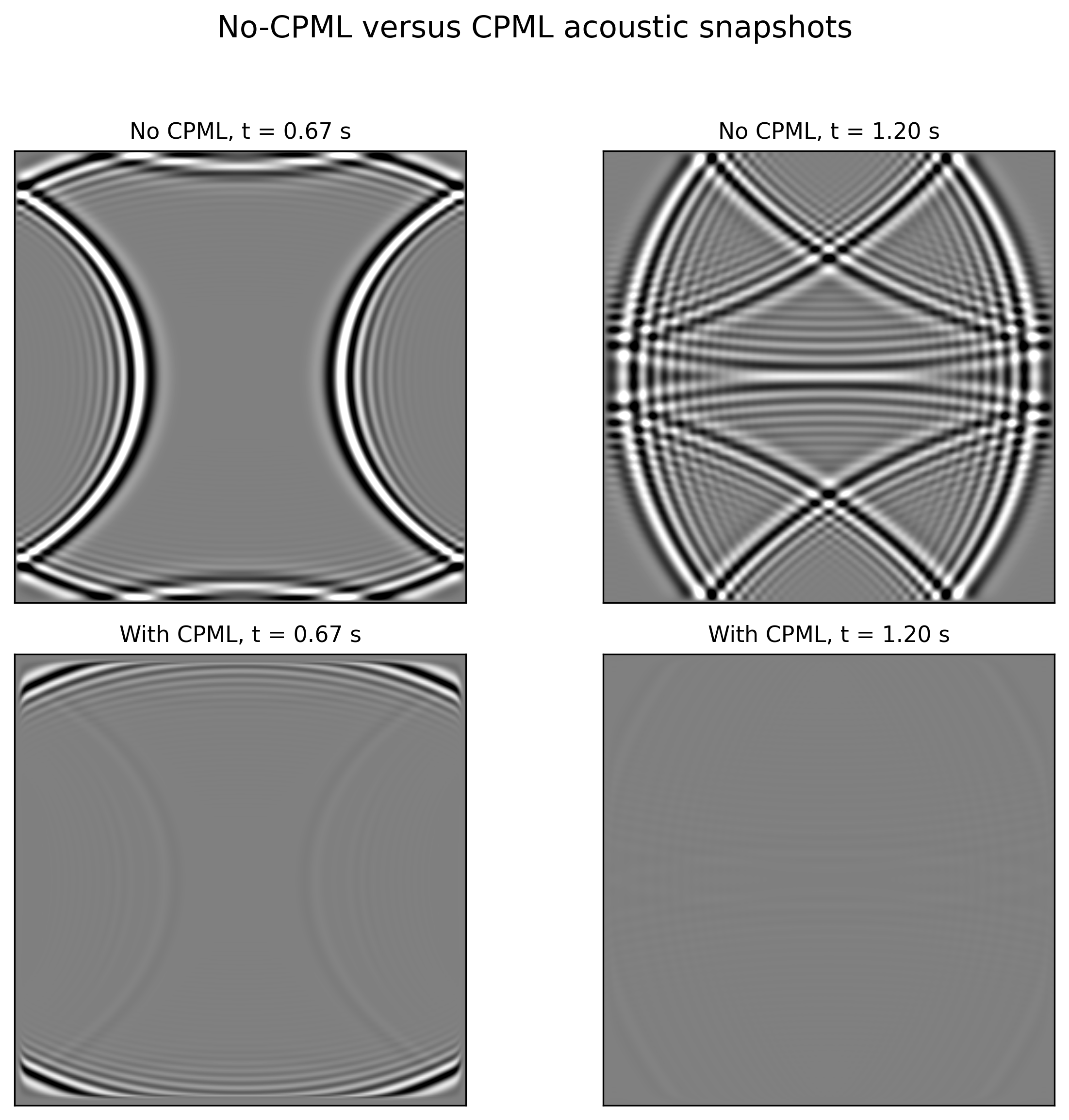}
    \caption{\bf Acoustic CPML wavefield visualization using a Ricker pulse
    source. The no-CPML case retains reflected waves, while the CPML case
    strongly attenuates the outgoing wave and reduces boundary reflections.}
    \label{fig:cpml_wavefield_snapshots}
\end{figure}

\begin{figure}[htbp]
    \centering
    \includegraphics[width=\columnwidth]{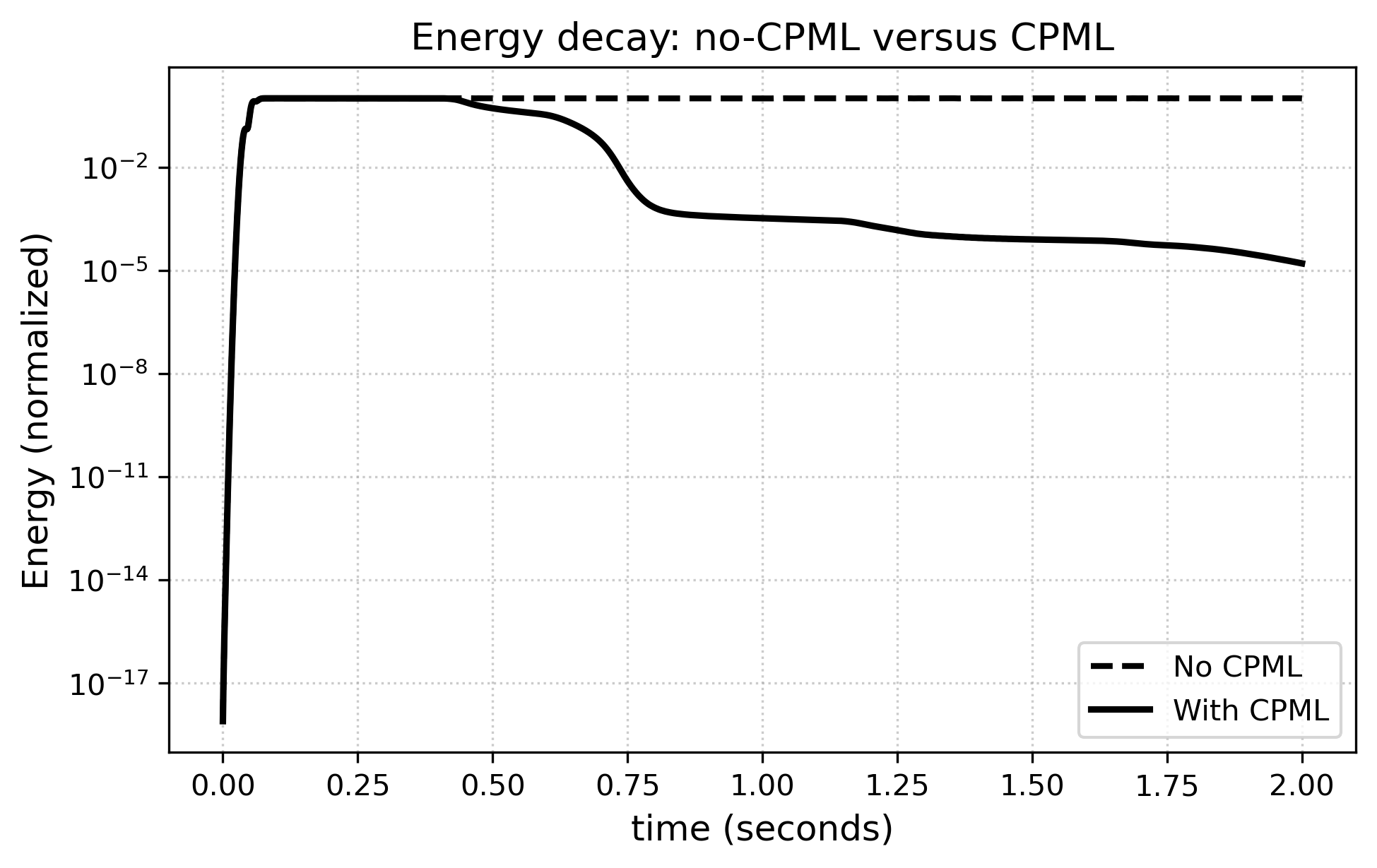}
    \caption{\bf Normalized field energy with and without CPML.  After the
    source injection ends, the no-CPML case retains energy due to boundary
    reflections, while the CPML case decays by several orders of magnitude as
    outgoing waves are absorbed.}
    \label{fig:cpml_energy_decay}
\end{figure}

Figure~\ref{fig:cpml_wavefield_snapshots} compares pressure snapshots with and
without CPML.  In the no-CPML case, waves reflect from the artificial boundary
and remain visible inside the computational domain.  In the CPML case, the
outgoing wave is strongly attenuated as it enters the absorbing layer, leaving
a much smaller reflected field.  Figure~\ref{fig:cpml_energy_decay} shows the
corresponding normalized field energy.  After the source injection ends, the
no-CPML simulation retains energy because waves are trapped by boundary
reflections, while the CPML simulation decays by several orders of magnitude as
outgoing waves are absorbed.

Figures~\ref{fig:cpml_wavefield_snapshots} and~\ref{fig:cpml_energy_decay}
show a separate single-domain acoustic CPML demonstration. They are included
only to illustrate the absorbing behavior of CPML visually, while
Table~\ref{tab:cpml_reflection_decomposition_validation} provides the
decomposition-dependent quantitative check used for the multi-GPU
implementation.

\section{Performance Metrics}

Runtime is the wall-clock time required to advance the full 3D grid for \(N_t\)
time steps. Throughput is reported as
\(\mathrm{Mpoints/s}=N_xN_yN_zN_t/(10^6T_{\mathrm{runtime}})\).
Strong-scaling efficiency on \(p\) GPUs is
\(E_p=T_1/(pT_p)\), and the enlarged-ghost speedup relative to \(s=1\) is
\(S_s=T_{s=1}/T_s\). CPML overhead is computed as
\((T_{\rm CPML}-T_{\rm no\;CPML})/T_{\rm no\;CPML}\times100\%\).
\begin{figure}[t]
    \centering
    \includegraphics[width=\columnwidth, trim=20 20 20 20, clip]{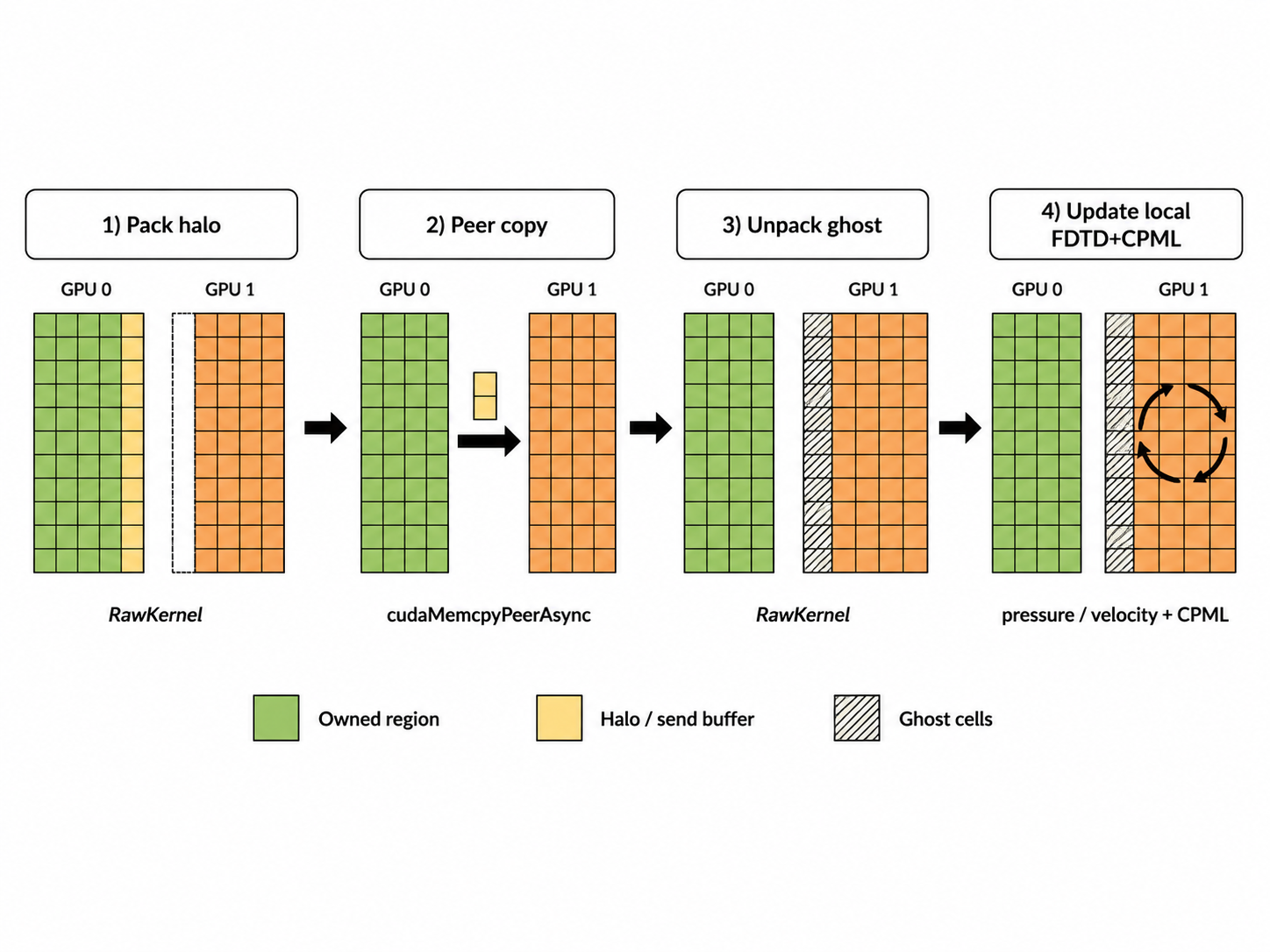}
    \caption{\bf Halo exchange workflow.}
    \label{fig:gpuhalo}
\end{figure}

\section{Results and Discussion}

\subsection{Decomposition Comparison}

Most experiments were performed on the four-GPU NVIDIA Quadro RTX 6000 node described in Table 3. 
The strong-scaling experiment in Table \ref{tab:rtx8000_strong_scaling_memory} was performed separately on NVIDIA Quadro RTX 8000 GPUs, and is reported separately because it uses different hardware.  Each entry gives runtime in seconds followed by throughput in
Mpoints/s.

\subsection{Raw-Kernel and Communication Optimizations}

The optimized implementation used in the performance experiments replaces
high-level array operations with explicit raw CUDA kernels launched through
CuPy.  Each kernel assigns CUDA threads directly to grid points in the local
subdomain, so the pressure field, velocity fields, and CPML auxiliary memory
variables are updated using contiguous device arrays.  This reduces Python-level
overhead and avoids repeated temporary-array creation during the time-stepping
loop.

For the multi-GPU runs, each GPU stores an owned subdomain together with ghost
cells along internal GPU interfaces.  Before each local update, halo data are
packed into contiguous device buffers using raw CUDA packing kernels.  These
buffers are then transferred directly between neighboring GPUs using CUDA
peer-to-peer copies, and unpacked into the receiving ghost cells using raw CUDA
unpacking kernels.  This avoids host-staged communication of the form
GPU--CPU--GPU and instead uses direct GPU--GPU exchange whenever peer access is
available.

The optimized code also reuses communication buffers rather than allocating new
buffers at every time step.  For a standard halo exchange, \(s=1\), the halo is
exchanged every time step.  For enlarged ghost-region communication, each GPU
exchanges a deeper ghost region of depth \(g=2rs\), where \(r\) is the stencil
radius and \(s\) is the number of local steps taken before the next exchange.
This reduces the number of halo exchanges by a factor of approximately \(s\),
but it also introduces redundant computation in the enlarged ghost region.

The performance tables below use this raw-kernel communication path for the
pencil-\(yz\) layout.  With the decomposition fixed, the experiments isolate the
effect of communication strategy: standard one-step halo exchange, enlarged
ghost regions, direct GPU-to-GPU peer exchange, and host-staged exchange.

\subsection{Correctness Check}

\begin{table}[t]
\centering
\caption{\bf Correctness check for enlarged ghost-region communication in the four-GPU pencil-\(yz\) first-order acoustic FDTD+CPML implementation. For each grid, the table reports the maximum relative difference over \(s\in\{2,4,8\}\) compared with the standard \(s=1\) halo-exchange reference solution.}
\label{tab:correctness_check}
\renewcommand{\arraystretch}{1.15}
\setlength{\tabcolsep}{5pt}
\resizebox{\columnwidth}{!}{%
\begin{tabular}{|c|c|c|c|}
\hline
\bf Grid & \bf GPUs & \bf Max relative \(L^2\) difference & \bf Max relative \(L^\infty\) difference \\
\hline
\(320^3\) & 4 & \(0.000\times 10^{0}\) & \(0.000\times 10^{0}\) \\
\hline
\(480^3\) & 4 & \(0.000\times 10^{0}\) & \(0.000\times 10^{0}\) \\
\hline
\(544^3\) & 4 & \(0.000\times 10^{0}\) & \(0.000\times 10^{0}\) \\
\hline
\(640^3\) & 4 & \(0.000\times 10^{0}\) & \(0.000\times 10^{0}\) \\
\hline
\(800^3\) & 4 & \(0.000\times 10^{0}\) & \(0.000\times 10^{0}\) \\
\hline
\end{tabular}
}
\end{table}

Table~\ref{tab:correctness_check} shows that the enlarged ghost-region
solutions agree with the standard \(s=1\) reference to the reported precision.
The zero values in Table~\ref{tab:correctness_check} mean that the differences are below the printed precision; in these tests the enlarged-ghost solutions were numerically indistinguishable from the standard \(s=1\) reference.

\subsection{Single-GPU Baseline}

\begin{table}[t]
\centering
\caption{\bf Single-GPU raw-kernel throughput for the first-order acoustic
3D FDTD benchmark with and without CPML boundary-layer updates.}
\label{tab:single_gpu_baseline}
\renewcommand{\arraystretch}{1.15}
\setlength{\tabcolsep}{4pt}
\resizebox{\columnwidth}{!}{%
\begin{tabular}{|c|c|c|c|c|}
\hline
\bf Grid & \bf Boundary & \bf Runtime (s) & \bf Mpoints/s & \bf CPML overhead \\
\hline
$320^3$ & none & 0.635 & 3304.48 & -- \\
\hline
$320^3$ & CPML & 0.638 & 3289.58 & 0.45\% \\
\hline
$480^3$ & none & 2.440 & 2900.31 & -- \\
\hline
$480^3$ & CPML & 2.449 & 2889.58 & 0.37\% \\
\hline
$544^3$ & none & 3.530 & 2918.65 & -- \\
\hline
$544^3$ & CPML & 3.553 & 2900.14 & 0.64\% \\
\hline
$640^3$ & none & 5.751 & 2917.52 & -- \\
\hline
$640^3$ & CPML & 5.766 & 2909.64 & 0.27\% \\
\hline
$800^3$ & none & 11.204 & 2924.73 & -- \\
\hline
$800^3$ & CPML & 11.221 & 2920.36 & 0.15\% \\
\hline
\end{tabular}
}
\end{table}

Table~\ref{tab:single_gpu_baseline} shows that the measured single-GPU CPML
boundary-layer overhead remains below \(1\%\) for all tested grids.  Since the
CPML update is confined to a thin boundary layer, its relative cost is small
compared with the full-volume FDTD update.  Small nonmonotone variations across
grid sizes are expected because the overhead is computed from small differences
between separate runtime measurements.

\subsection{Strong Scaling and Memory Capacity on RTX 8000 GPUs}

\begin{table}[t]
\centering
\caption{\bf Strong-scaling and memory-capacity behavior of the raw-kernel
first-order acoustic FDTD+CPML benchmark on NVIDIA Quadro RTX 8000 GPUs using
standard \(s=1\) halo exchange.}
\label{tab:rtx8000_strong_scaling_memory}
\renewcommand{\arraystretch}{1.15}
\setlength{\tabcolsep}{4pt}
\resizebox{\columnwidth}{!}{%
\begin{tabular}{|c|c|c|c|c|}
\hline
\bf Grid & \bf GPUs & \bf Runtime (s) & \bf Speedup & \bf Parallel efficiency \\
\hline
\multirow{3}{*}{$320^3$}
& 1 & 0.897 & 1.00 & 100.0\% \\ \cline{2-5}
& 2 & \bf 0.864 & \bf 1.04 & \bf 51.9\% \\ \cline{2-5}
& 4 & 1.600 & 0.56 & 14.0\% \\
\hline
\multirow{3}{*}{$544^3$}
& 1 & 4.666 & 1.00 & 100.0\% \\ \cline{2-5}
& 2 & \bf 3.442 & \bf 1.36 & \bf 67.8\% \\ \cline{2-5}
& 4 & 4.482 & 1.04 & 26.0\% \\
\hline
\multirow{3}{*}{$800^3$}
& 1 & 14.614 & 1.00 & 100.0\% \\ \cline{2-5}
& 2 & \bf 9.658 & \bf 1.51 & \bf 75.7\% \\ \cline{2-5}
& 4 & 10.458 & 1.40 & 34.9\% \\
\hline
\multirow{3}{*}{$1024^3$}
& 1 & OOM & -- & -- \\ \cline{2-5}
& 2 & 19.050 & -- & -- \\ \cline{2-5}
& 4 & \bf 18.528 & -- & -- \\
\hline
\end{tabular}
}
\end{table}

Table~\ref{tab:rtx8000_strong_scaling_memory} shows two regimes.  For grids up
to \(800^3\), the best runtime is obtained with two GPUs; four GPUs still gives
speedup over one GPU for the \(544^3\) and \(800^3\) cases, but it is slower
than the two-GPU configuration because the additional halo-exchange and
synchronization costs outweigh the extra reduction in local work.  For larger
grids, the memory-capacity benefit becomes more important.  The \(1024^3\) case
runs slightly faster on four GPUs than on two GPUs.
Thus, additional GPUs become useful not only for runtime reduction, but also
for enabling larger CPML simulations that do not fit on fewer devices.

\subsection{Enlarged Ghost-Region Performance}
After selecting the pencil-\(yz\) layout from the baseline decomposition study in Table~\ref{tab:decomposition_comparison}, we apply the optimized raw-kernel communication implementation and study the effects of enlarged ghost regions and direct GPU-to-GPU peer exchange.

\begin{table}[t]
\centering
\caption{\bf Four-GPU enlarged ghost-region performance for the optimized
raw-kernel pencil-\(yz\) first-order acoustic FDTD+CPML benchmark. The case
\(s=1\) is the standard one-step halo-exchange method.}
\label{tab:enlarged_ghost_performance}
\renewcommand{\arraystretch}{1.15}
\setlength{\tabcolsep}{4pt}
\resizebox{\columnwidth}{!}{%
\begin{tabular}{|c|c|c|c|c|c|}
\hline
\bf Grid & \bf GPUs & \bf $s$ & \bf Runtime (s) & \bf Mpoints/s & \bf Speedup vs. $s=1$ \\
\hline
\multirow{4}{*}{$320^3$}
& 4 & 1 & 1.708 & 1227.87 & 1.00 \\ \cline{2-6}
& 4 & 2 & 1.556 & 1348.07 & 1.10 \\ \cline{2-6}
& 4 & 4 & 1.491 & 1406.48 & 1.15 \\ \cline{2-6}
& 4 & 8 & 1.568 & 1337.26 & 1.09 \\
\hline
\multirow{4}{*}{$480^3$}
& 4 & 1 & 3.669 & 1929.27 & 1.00 \\ \cline{2-6}
& 4 & 2 & 3.521 & 2010.15 & 1.04 \\ \cline{2-6}
& 4 & 4 & 3.448 & 2052.46 & 1.06 \\ \cline{2-6}
& 4 & 8 & 3.636 & 1946.61 & 1.01 \\
\hline
\multirow{4}{*}{$544^3$}
& 4 & 1 & 4.791 & 2150.36 & 1.00 \\ \cline{2-6}
& 4 & 2 & 4.599 & 2240.38 & 1.04 \\ \cline{2-6}
& 4 & 4 & 4.514 & 2282.74 & 1.06 \\ \cline{2-6}
& 4 & 8 & 4.727 & 2179.67 & 1.01 \\
\hline
\multirow{4}{*}{$640^3$}
& 4 & 1 & 6.781 & 2474.16 & 1.00 \\ \cline{2-6}
& 4 & 2 & 6.537 & 2566.60 & 1.04 \\ \cline{2-6}
& 4 & 4 & 6.397 & 2622.52 & 1.06 \\ \cline{2-6}
& 4 & 8 & 6.654 & 2521.52 & 1.02 \\
\hline
\multirow{4}{*}{$800^3$}
& 4 & 1 & 11.100 & 2951.97 & 1.00 \\ \cline{2-6}
& 4 & 2 & 10.738 & 3051.65 & 1.03 \\ \cline{2-6}
& 4 & 4 & 10.444 & 3137.44 & 1.06 \\ \cline{2-6}
& 4 & 8 & 10.799 & 3034.24 & 1.03 \\
\hline
\end{tabular}
}
\end{table}

Table~\ref{tab:enlarged_ghost_performance} shows that enlarged ghost exchange
improves performance for all tested grids,with the best performance at \(s=4\). The \(s=8\) case is slower than \(s=4\), indicating that excessive
ghost enlargement introduces enough redundant work to reduce the benefit of
fewer exchanges.

\subsection{Host-Staged Versus Peer Exchange}

\begin{table}[t]
\centering
\caption{\bf Comparison of host-staged and direct GPU-to-GPU peer exchange for the four-GPU raw-kernel pencil-\(yz\) first-order acoustic FDTD+CPML benchmark. All runs use the standard \(s=1\) halo-exchange method.}
\label{tab:host_vs_peer_exchange}
\renewcommand{\arraystretch}{1.15}
\setlength{\tabcolsep}{4pt}
\resizebox{\columnwidth}{!}{%
\begin{tabular}{|c|c|c|c|c|c|}
\hline
\bf Grid & \bf GPUs & \bf Exchange method & \bf Runtime (s) & \bf Mpoints/s & \bf Speedup \\
\hline
\multirow{2}{*}{$320^3$}
& 4 & Host-staged & 4.720 & 444.33 & 1.00 \\ \cline{2-6}
& 4 & GPU-to-GPU peer & 1.708 & 1227.87 & \bf 2.76 \\
\hline
\multirow{2}{*}{$480^3$}
& 4 & Host-staged & 9.628 & 735.17 & 1.00 \\ \cline{2-6}
& 4 & GPU-to-GPU peer & 3.669 & 1929.27 & \bf 2.62 \\
\hline
\multirow{2}{*}{$544^3$}
& 4 & Host-staged & 12.223 & 842.97 & 1.00 \\ \cline{2-6}
& 4 & GPU-to-GPU peer & 4.791 & 2150.36 & \bf 2.55 \\
\hline
\multirow{2}{*}{$640^3$}
& 4 & Host-staged & 16.853 & 995.48 & 1.00 \\ \cline{2-6}
& 4 & GPU-to-GPU peer & 6.781 & 2474.16 & \bf 2.49 \\
\hline
\multirow{2}{*}{$800^3$}
& 4 & Host-staged & 27.278 & 1201.24 & 1.00 \\ \cline{2-6}
& 4 & GPU-to-GPU peer & 11.100 & 2951.97 & \bf 2.46 \\
\hline
\end{tabular}
}
\end{table}

Table~\ref{tab:host_vs_peer_exchange} shows that direct GPU-to-GPU peer
exchange gives \textbf{\(2.46\times\)--\(2.76\times\) }speedup over host-staged exchange.

\subsection{Memory Capacity Benefit}

\begin{table}[t]
\centering
\caption{\bf Estimated per-GPU memory footprint for the raw-kernel first-order acoustic FDTD+CPML implementation using standard \(s=1\) halo exchange. The estimate includes the pressure field, three velocity fields, and six CPML auxiliary memory fields in single precision.}
\label{tab:memory_capacity_benefit}
\renewcommand{\arraystretch}{1.15}
\setlength{\tabcolsep}{4pt}
\resizebox{\columnwidth}{!}{%
\begin{tabular}{|c|c|c|c|c|}
\hline
\bf Grid & \bf GPUs & \bf Decomposition & \bf Estimated memory/GPU & \bf Memory reduction \\
\hline
\multirow{3}{*}{$320^3$}
& 1 & $1\times1\times1$ & 1.41 GiB & 1.00$\times$ \\ \cline{2-5}
& 2 & $1\times1\times2$ & 0.74 GiB & \textbf{1.91}$\times$ \\ \cline{2-5}
& 4 & $1\times2\times2$ & 0.39 GiB & \textbf{3.64}$\times$ \\
\hline
\multirow{3}{*}{$544^3$}
& 1 & $1\times1\times1$ & 6.54 GiB & 1.00$\times$ \\ \cline{2-5}
& 2 & $1\times1\times2$ & 3.36 GiB & \textbf{1.94}$\times$ \\ \cline{2-5}
& 4 & $1\times2\times2$ & 1.73 GiB & \textbf{3.78}$\times$ \\
\hline
\multirow{3}{*}{$800^3$}
& 1 & $1\times1\times1$ & 20.24 GiB & 1.00$\times$ \\ \cline{2-5}
& 2 & $1\times1\times2$ & 10.32 GiB &\textbf{ 1.96}$\times$ \\ \cline{2-5}
& 4 & $1\times2\times2$ & 5.26 GiB & \textbf{3.85}$\times$ \\
\hline
\end{tabular}
}
\end{table}

Table~\ref{tab:memory_capacity_benefit} reports an analytical estimate of the
per-GPU field storage.  For grids that fit comfortably on one GPU, single-GPU
execution can be more cost-effective because it avoids communication overhead.
For larger CPML simulations, however, multi-GPU decomposition reduces the
per-GPU memory footprint and enables runs that are close to or beyond the
capacity of one device.

\section{Conclusions and Future Work}

Overall, the experiments show that the main benefit of the multi-GPU
implementation is not universal strong-scaling speedup over a highly
optimized single-GPU solver. Rather, the main benefits are communication
efficiency and memory scalability. Direct GPU-to-GPU peer exchange removes
most of the host-staging bottleneck, while multi-GPU decomposition reduces
the per-GPU memory footprint and enables larger CPML simulations. Enlarged
ghost regions can further reduce communication frequency, but their benefit
is limited by redundant computation and extra memory traffic.

Future work will extend the implementation to multi-node GPU systems with improved communication overlap using CUDA streams, NCCL, or GPU-aware MPI. A second direction is to extend the study to full Maxwell systems, heterogeneous media, higher-order CPML models, and other boundary conditions, with the goal of automatically selecting decomposition layouts and ghost-region depths for a given hardware topology and problem size. Another direction is to combine the present multi-GPU FDTD+CPML implementation
with learned structure-preserving spatial discretizations, such as
energy-conserving data-driven convolution stencils for Maxwell-type systems
\cite{obieke2026energystableapproachlearning}.


\section*{Availability of Data and Code}
The code and scripts used to generate the numerical and performance results in
this study are available at
\url{https://github.com/victoryobieke/fdtd-cpml-multigpu}.

\section*{Conflict of Interest}
The author declares that there are no conflicts of interest.

\bibliographystyle{abbrv}
\bibliography{references}

\end{document}